\documentclass{ws-procs9x6}

\usepackage{amssymb,bbold}
\usepackage{hyperref}

\newcommand{\be}[3]{\begin{equation}  \label{#1#2#3}}
\newcommand{\ee}{\end{equation}}
\newcommand{\ba}{\begin{array}}
\newcommand{\ea}{\end{array}}
\newcommand{\bea}[3]{\begin{eqnarray}  \label{#1#2#3}}
\newcommand{\eea}{\end{eqnarray}} 

\def\1{\mathbb 1}
\def\F{{\mathcal F}^{(1)}}
\def\FF{{\mathcal F}^{(7)}}
\def\FFF{{\mathcal F}^{(27)}}
\def\N{\mathcal N}

\newcommand{\haken}{\mathbin{\hbox to 8pt{%
                 \vrule height0.4pt width7pt depth0pt
                 \kern-.4pt
                 \vrule height4pt width0.4pt depth0pt\hss}}}

\begin{document}

\title{
Fluxes in M-theory on 7-manifolds: \\[3mm]
$G_2$-, SU(3)- and SU(2)-structures}

\author{K. Behrndt\footnote{\uppercase{W}ork partially
supported by a \uppercase{H}eisenberg grant of the \uppercase{DFG}}}

\address{ Max-Planck-Institut f\"ur Gravitationsphysik,
Albert Einstein Institut\\
Am M\"uhlenberg 1,  14476 Golm, Germany \\
E-mail: behrndt@aei.mpg.de}

\author{C. Jeschek\footnote{\uppercase{W}ork supported by a
\uppercase{G}raduiertenkolleg grant of the \uppercase{DFG}
(\uppercase{T}he \uppercase{S}tandard \uppercase{M}odel of
\uppercase{P}article \uppercase{P}hysics - structure, precision tests
and extensions). }}

\address{ Humboldt Universit\"at zu Berlin,
Institut f\"ur Physik,\\
Newtonstrasse 15, 12489 Berlin,
 Germany\\
E-mail: jeschek@physik.hu-berlin.de
}  

\maketitle

\abstracts{We consider compactifications of M-theory on 7-manifolds in
the presence of 4-form fluxes, which leave at least four supercharges
unbroken. Supersymmetric vacua admit $G$-structures and we discuss the
cases of $G_2$-, SU(3)- as well as SU(2)-structures.  We derive the
constraints on the fluxes imposed by supersymmetry and determine the
flux components that fix the resulting 4-dimensional cosmological
constant (i.e.\ superpotential).}

\vspace{2cm}

\begin{center}
{\em To appear in Proceedings of: BW2003 Workshop, \\
29 Aug. - 02 Sept., 2003 Vrnjacka Banja, Serbia
}
\end{center}


\newpage


\section{Introduction}


An essential input in lifting the continuous moduli space might be
non-zero fluxes on the internal space.  By now one can find a long
list of literature about this subject.  A starting point was the work
by Candelas and Raine\cite{480} for an un-warped metric which was
generalized later in\cite{490} (for an earlier work on warp
compactification see \cite{500}) and the first examples, which
preserve $\N = 1$ supersymmetry appeared in\cite{520}. The subject
was revived around 10 years later by the work of Polchinski and
Strominger\cite{300}, where flux compactifications in type II string
theory was considered. In the M-theory setting, different aspects are
discussed in\cite{460,270,120,240,130,110,470,400,640}.

Fluxes induces a non-trivial back reaction onto the geometry, because
they appear as specific con-torsion
components\cite{530,570,250,350,320,240,170,470,660,650} for the
Killing spinor. The resulting spaces are in general non-K\"ahlerian,
which reflects the fact that the moduli space is (partly) lifted.  In
order to see which moduli are fixed, one can derive the corresponding
superpotential as function of the fluxes in a way discussed
in\cite{160}, but this approach becomes subtle if the fluxes are not
related to closed forms (due to Chern-Simons terms).

In this talk we discuss M-theory compactifications in the presence of
4-form fluxes, which keep the external 4-d space time maximal
symmetric, i.e.\ either flat or anti deSitter (AdS), where in the
latter case the superpotential remains non-zero in the vacuum giving
rise to a negative cosmological constant. We start by making the
Ansatz for the metric and the 4-form field strength and separate the
gravitino variation into an internal and external part. In addition,
we have to make an Ansatz for the 11-d Killing spinor, which
decomposes into internal 7-d spinors and the external 4-d spinors. In
the most general case, the solution will be rather involved and we use
$G$-structures to classify possible vacua (Section 3).  These
structures are defined by a set of invariant differential forms and
are in one to one correspondence to the number of internal spinors,
which will enter the 11-d Killing spinor.  Using these differential
forms, one can formely solve the BPS equations (Section 4), but
explicit solutions require the construction of these forms.  Note, the
case of the $G_2$- and SU(3)-structures have been discussed already
before\cite{130,110,470} and we will be rather short.


\section{Warp compactification in the presence of fluxes}


In the (flux) vacuum, all Kaluza-Klein scalars and vectors are trivial
and hence we consider as Ansatz for the metric and the 4-form field
strength
\be012
\ba{rcl}
ds^2 &=& e^{2 A} \big[ \, g^{(4)}_{\mu\nu} dx^\mu dx^\nu
        + \, h_{ab} dy^a dy^b \big] \ , \\[2mm]
\hat F &=& \frac{m}{4!} \,  
\epsilon_{\mu\nu\rho\lambda} dx^\mu \wedge dx^\nu \wedge
    dx^\rho \wedge dx^\lambda +
\frac{1}{4!} F_{abcd} \, dy^a \wedge dy^b \wedge dy^c \wedge dy^d
\ea
\ee
where $A=A(y)$ is a function of the coordinates of the 7-manifold with
the metric $h_{ab}$, $m$ is the Freud-Rubin parameter and the 4-d
metric $g^{(4)}_{\mu\nu}$ is either flat or anti deSitter.  Unbroken
supersymmetry requires the existence of (at least) one Killing spinor
$\eta$ yielding a vanishing gravitino variation of 11-dimensional
supergravity
\be716
\ba{rcl}
0 = \delta \Psi_M &=& 
        \big[ \partial_M + {1 \over 4} \hat \omega^{RS}_M \Gamma_{RS}
        + {1 \over 144} \big(\Gamma_M  F - 12 \,  F_M \big) \big] \eta 
\ea
\ee
where: $\hat F \equiv F_{MNPQ} \Gamma^{MNPQ}$ , $\hat F_M \equiv
F_{MNPQ} \Gamma^{NPQ}$, etc.  Since,
\be371
\Gamma_M \Gamma^{N_1 \cdots N_n} = \Gamma_M^{\ N_1 \cdots N_n} +
n \, \delta_M^{\ [N_1} \Gamma^{N_2 \cdots N_n]}
\ee
one can bring the variation also in the more common form.  Using the
convention $ \{ \Gamma^A , \Gamma^B \} = 2 \eta^{AB}$ with $\eta =
{\rm diag}(-,+,+ \ldots +)$, we decompose the $\Gamma$-matrices as
usual
\be391
\Gamma^\mu = \hat \gamma^\mu \otimes \1 \qquad , \qquad
\Gamma^{a+3} = \hat \gamma^5 \otimes \gamma^a
\ee
with $\mu = 0..3$, $a = 1..7$, and $\hat \gamma^5 = i \hat
\gamma^0 \hat \gamma^1 \hat \gamma^2 \hat \gamma^3$, \ $1=i \gamma^1
\gamma^2 \gamma^3 \gamma^4 \gamma^5 \gamma^6 \gamma^{7}$ yields
\be161
i \hat \gamma^5 \hat \gamma^\mu = {1 \over 3!} \epsilon^{\mu\nu\rho\lambda}
\hat \gamma_{\nu\rho\lambda} \quad ,\qquad
{i \over 3!} \epsilon^{abcdmnp} \gamma_{mnp} =
\gamma^{abcd} \equiv \gamma^{[a} \gamma^b \gamma^c \gamma^{d]} \ .
\ee
The spinors in 11-d supergravity are Majorana and we take all 4-d
$\hat \gamma^\mu$-matrices are real and $\hat \gamma^5$ as well as the
7-d $\gamma^a$-matrices are purely imaginary and antisymmetric.

With this notation, we can now split the gravitino variation into an
internal and external part. In order to deal with the warp factor, we
use
\be625
ds^2 = e^{2A} \widetilde{ds}^2 \quad \rightarrow \quad
D_M = \tilde D_M + {1 \over 2}    \Gamma_M^{\ N} \partial_N A
\ee
and find for the external components of the gravitino variation
\be112
0=\big[ \nabla_\mu \otimes \1 +  \hat \gamma_\mu \hat \gamma^5
  \otimes \big( {1 \over 2} \, \partial A + {i \, m \over 36} \big)
+ {1 \over 144} e^{-3A} \, \hat \gamma_\mu \otimes F
  \big] \eta  
\ee
where $F=F_{abcd}\gamma^{abcd}$, $F_a = F_{abcd}\gamma^{bcd}$, etc.\
and $\nabla_\mu$ is the 4-d covariant derivative.  In the same way, we
get for the internal variation
\be711
0= \big[ \1 \otimes
\big( \nabla_a^{(h)} -  {1 \over 2} \partial_a A  
+ { i m \over 48} \gamma_a \big)  - 
{1\over 4} \hat \gamma^5  \hat \gamma^\mu \nabla_\mu \otimes \gamma^a -
{1 \over 12}\, e^{-3A}\, \hat \gamma^5 \otimes F_a  \big] \eta 
\ee
where we eliminated the term $\sim \gamma_a F \eta$ by using eq.\
(\ref{112}).

In order to solve these equations, we have to decompose the spinor and
introduce the superpotential yielding the negative cosmological
constant.  The 11-d Majorana spinor can be expanded in all independent
spinors as
\[
\eta = \sum_{i=1}^\N (\epsilon^i \otimes \theta^i + cc) \ ,
\]
where $\epsilon^i$ and $\theta^i$ denote the 4- and 7-d spinors, resp.
If there are no fluxes, all of these spinors are covariantly constant
and $\N \leq 8$ gives the resulting extended supersymmetries in 4
dimensions.  With non-trivial fluxes one can however impose a relation
between the spinors and $\N$ does not refer to the number of unbroken
supersymmetries (see last Section), but gives nevertheless a
classification of supersymmetric vacua. In fact, with these spinors
one can build differential forms that are singlets under a subgroup $G
\subset spin(7)$ and hence define a $G$-structure, where the number of
spinors is directly related to the group $G$ (see next Section).
By definition, the spinors are singlets under $G$ and therefore obey
certain projector conditions, which annihilate all non-singlet
components and, at the same time, can be used to derive simple
differential equations for the spinors and constraints on the fluxes
(see last Section).

If the 4-d spinors are covariantly constant, the resulting vacuum will
be a 4-d flat space, but for an anti deSitter vacuum
the spinors satisfy 
\be811
\nabla_\mu \epsilon^i \  \sim \ \hat \gamma_\mu \, 
 ( W_1^{ij} + i \hat \gamma^5 \,W_2^{ij} ) 
\, \epsilon_j \ .
\ee
Note, the resulting 4-d cosmological constant will be: $- |W|^2$ and
we did not take into account a K\"ahler potential, ie.\ our
superpotential will not be holomorphic.  If there is only a single
spinor this equation simplifies to
\[
\nabla_\mu \epsilon ~ \sim ~ 
\hat \gamma_\mu  \, (W_1 + i \, \hat \gamma^5 \, W_2 ) \, \epsilon 
\]
and if $\epsilon$ is a Weyl spinor it becomes $\nabla_\mu \epsilon =
\hat \gamma_\mu  \, \bar W \epsilon^\star$ with the complex
superpotential $W = W_1 +i\, W_2$. If $\epsilon_i$ are a set of Weyl
spinors, we introduce the superpotential by a 11-d spinor satisfying
the equation
\be152
\big[\nabla_\mu \otimes \1 \big] \eta = (\hat \gamma_\mu \otimes 
\1) \tilde \eta
\qquad {\rm with:} \qquad
\tilde \eta =  W^{ij} \epsilon_i \otimes \theta^\star_j
+ cc  
\ee
This way of introducing the superpotential might be confusing.
Recall, we set constant all 4-d scalars as well as vector potentials
and hence the superpotential should just be a number fixing the
cosmological constant for the given vacuum. Since we introduced the
superpotential in the 11-d Killing spinor equation it will, on the
other hand, depend on the fluxes and the warp factor and thus it is in
general not constant over the internal space.  The correct
4-dimensional superpotential is of course obtained only after a
Kaluza-Klein reduction, i.e.\ after an integration over the internal
space and to make this clear we will denote this constant
superpotential by $W^{(0)}$.  We do not want to discuss issues related
to a concrete Kaluza-Klein reduction (over a not Ricci-flat internal
space) and want instead determine the flux components that are
responsible for a non-zero value of $W^{(0)}$.


\section{$G$-Structures}


Supersymmetric compactifications on 7-manifolds imply the existence of
differential forms, which are singlets under a group $G \subset
spin(7)$ and which define $G$-structures\footnote{We follow here basically the
procedure initiated in the recent discussion by\cite{250}.}.  These
globally defined differential forms can be constructed as bi-linears
of the internal Killing spinors as
\[
\theta_i \gamma_{a_1 \cdots a_n} \theta_j 
\]
and the group $G$ is fixed by the number of independent spinors
$\theta_i$ which are all singlets under $G$.  E.g.\ if there is only a
single spinor on the 7-manifold, it can be chosen as a real $G_2$
singlet; if there are two spinors, one can combine them into a complex
$SU(3)$ singlet; three spinors can be written as $Sp(2)\simeq SO(5)$
singlets and four spinors as $SU(2)$ singlets.  Of course, all eight
spinors cannot be a singlet of a non-trivial subgroup of $SO(7)$ and
$G$ is trivial. The 7-dimensional $\gamma$-matrices are in the
Majorana representation and satisfy the relation: $(\gamma_{a_1 \cdots
a_n})^T =(-)^{n^2 +n \over 2}\gamma_{a_1 \cdots a_n}$, which implies
that the differential form is antisymmetric in $[i,j]$ if $n=1,2,5,6$
and otherwise symmetric [we assumed here of course that $\theta^i$ are
commuting spinors and the external spinors are anti-commuting].  This
gives the well-known statement that having only a single spinor, one
cannot build a vector or a 2-form, but only a 3-form and its dual
4-form [the 0- and 7-form exist trivially on any spin manifold]. If we
have two spinors $\theta_{\{1/2\}}$, we can build one vector $v$ and
one 2-form (and of course its dual 5- and 6-form). Since the spinors are
globally well-defined, also the vector field is well defined on $X_7$
and it can be used to obtain a foliation of the 7-d space by a
6-manifold $X_6$.  Similarly, having three 7-spinors we can build
three vector fields as well as three 2-forms and having four spinors
the counting yields six vectors combined with six 2-forms. In addition
to these vector fields and 2-forms, one obtains further 3-forms the
symmetrized combination of the fermionic bi-linears. We have however
to keep in mind, that all these forms are not independent, since Fierz
re-arrangements yield relations between the different forms,
see\cite{250,240} for more details.

Using complex notation, we can introduce the following two sets of bi-linears 
[$\hat{\theta}^{\dagger}=(\hat{\theta}^{\ast})^T$]:
\[
\Omega_{a_1 \cdots a_k } \, \equiv \, 
     {\theta}^{\dagger}\gamma_{a_1 \cdots a_k} {\theta}
\qquad \mbox{and}\qquad
\tilde \Omega_{a_1 \cdots a_k} \, \equiv \, 
     {\theta}^T\gamma_{a_1 \cdots a_k} {\theta}
\]
where dropped the index $i,j$ which counts the spinors.
The associated $k$-forms becomes now
\be624
\Omega^k \, \equiv \, 
     {1 \over k!} \Omega_{a_1 \cdots a_k} e^{a_1 \cdots a_k}
\qquad \mbox{and}\qquad
\tilde \Omega^k \, \equiv \, 
     {1 \over k!} \tilde\Omega_{a_1 \cdots a_k} e^{a_1 \cdots a_k} \, .
\ee
If the spinors are covariantly constant (with respect to the
Levi-Civita connection) the group $G$ coincides with the holonomy of
the manifold. If the spinors are not covariantly constant neither can
be these differential forms and the deviation of $G$ from the holonomy
group is measured by the intrinsic torsion.  In the following we will
discuss the different cases in more detail.


\subsection*{$G_2$ structures}


In the simplest case, the Killing spinor is a $G_2$ singlet and reads
\be653
\theta  = e^{Z} \theta_0
\ee
where $\theta_0^T$ is a normalized real spinor.  Due to the properties
of the 7-d $\gamma$-matrices (yielding especially $\theta_0^T \gamma_a
\theta_0 =0$), only the following differential forms are non-zero
\be622
\ba{rcl}
1&=& \theta_0^T \theta_0 \ , \\[2mm]
i\, \varphi_{abc}&=&  \theta_0^T \gamma_{abc} \theta_0 \ , \\[2mm]
- \psi_{abcd} &=&  \theta_0^T \gamma_{abcd} \theta_0   \ , \\[2mm]
i\, \epsilon_{abcdmnp}& =&  \theta_0^T\gamma_{abcdmnp} \theta_0  \ .
\ea
\ee
They are  $G_2$-invariant since $\theta_0$ is a $G_2$ singlet, i.e.\ 
it obeys the appropriate projector constraints.  Note, the Lie algebra
$\mathfrak{so}(7)$ is isomorphic to $\Lambda^2$ and a reduction of the
structure group on a general $X_7$ from $SO(7)$ to the subgroup $G_2$
implies the following splitting:
\be980
\mathfrak{so}(7) \, = \, \mathfrak{g}_2 \, \oplus \, \mathfrak{g}_2^{\bot}
\ .
\ee
This induces a decomposition of the space of 2-forms in
the following irreducible $G_2$-modules, 
\be981
\Lambda^2 \, = \, \Lambda^2_7 \, \oplus \, \Lambda^2_{14} \, ,
\ee
where
\[
\ba{rcl}
\Lambda^2_7 & = & \{u\haken\varphi | u\in TX_7 \} 
    =  \{ \alpha \in \Lambda^2 \, | 
           \ast(\varphi \wedge \alpha)-2\alpha=0 \} \\[2mm]
\Lambda^2_{14} & = & \{ \alpha \in \Lambda^2 \, | 
           \ast(\varphi \wedge \alpha)+\alpha=0 \} \cong \mathfrak{g}_2
\ea
\]
with the abbreviation $u \haken \varphi \equiv u^{m}
\varphi_{mnp}$ and $\varphi$ denotes the $G_2$-invariant 3-index
tensor, which is expressed as fermionic bi-linear in (\ref{622}).  The
operator $\ast(\varphi \wedge \alpha)$ splits the 2-forms
correspondingly to the eigenvalues $2$ and $-1$. These relations
serve us to define the orthogonal projections $\mathcal{P}_k$ onto the
$k$-dimensional spaces:
\bea930
\mathcal{P}_7(\alpha) & = & {1 \over 3} \, 
              (\alpha + \ast(\varphi \wedge \alpha)) 
    =  {1 \over 3} \, ( \alpha + {1 \over 2} \alpha\haken\psi) \ ,\\
\mathcal{P}_{14}(\alpha)&  = & {1 \over 3} \, 
              (2\alpha - \ast(\varphi \wedge \alpha))
   =  {2 \over 3} \, ( \alpha - {1 \over 4} \alpha\haken\psi) 
\eea
where $\psi = \ast \varphi$.  To be concrete, the $G_2$-singlet spinor
satisfies the condition

\[
({\mathcal P}_{14})_{ \ ab}^{cd}\, \gamma_{cd} \, \theta_0 =
{2 \over 3} \big( \1^{cd}_{\ \ ab} - {1 \over 4} \psi^{cd}_{\ \ ab} \big) 
\gamma_{cd} \, \theta_0 =0 
\]
which is equivalent to
\be512 
\ba{rcl}
\gamma_{ab} \theta_0 &=& i \varphi_{abc} \gamma^c \theta_0 \ , \\[1mm]
\gamma_{abc} \theta_0 &=& \big( i
\varphi_{abc} + \psi_{abcd} \gamma^d \big) \, \theta_0 \ , \\[1mm]
\gamma_{abcd} \theta_0 &=& \big( - \psi_{abcd} - 4 i \varphi_{[abc}
\gamma_{d]} \big) \theta_0 
\ea
\ee
where the second and third conditions follow from the first one.
These relations can now be used to re-cast the Killing spinor
equations into constraints for the fluxes and differential equations
for the warp factor as well as the spinor $\theta$. In the generic
situation this spinor is not covariantly constant, which reflects the
fact that fluxes deform the geometry by the gravitational back
reaction. This can be made explicit by re-writing the flux terms as
con-torsion terms\footnote{There is also an ongoing discussion in the
mathematical literature, see\cite{370}.}
\[
\tilde \nabla_a \theta \equiv ( \nabla_a - {1 \over 4} \tau^{bc}_a
\gamma_{bc} ) \theta = 0 \ .
\]
{From} the symmetry it follows that $\tau$ has $7 \times 21 = 7 \times
(7+14)$ components, but if $\theta$ is a $G_2$-singlet the ${\bf 14}$
drops out and hence $\tau \in \Lambda^1 \otimes {\mathfrak
g}_2^{\perp}$. These components decompose under $G_2$ as
\[
{\bf 49} = {\bf 1 + 7 + 14 +27} = \tau_1 + \tau_7 + \tau_{14} + \tau_{27}
\]
where $\tau_i$ are called $G_2$-structures.  Since the Killing spinors
define $\varphi$ and $\psi$, these torsion classes can be obtained
from $d \varphi$ and $d \psi$ as follows
\be093
\ba{rcl}
d\varphi & \in & \Lambda^4 \, = \, \Lambda^4_1 \, \oplus \, \Lambda^4_7
\, \oplus \, \Lambda^4_{27}  \ , \\[2mm]
d \psi & \in & \, \Lambda^5 \, = \, \Lambda^5_7 \, \oplus \, 
\Lambda^5_{14}  \ ,
\ea
\ee
where the ${\bf 7}$ in $\Lambda^4_7$ is the same as in $\Lambda^5_7$
up to a multiple.  For a general 4-form $\beta$, the different
projections are
\be971
\ba{rcl}
\mathcal{P}_{1}(\beta) &=&{1 \over 4!}
\psi \haken \beta \ , \\[2mm]
\mathcal{P}^4_{7}(\beta) &=&-{1 \over 3!}  \varphi \haken \beta \ , \\[2mm]
\mathcal{P}_{27}(\beta)_{ab} &=& {1 \over 3!}(\beta_{cde\{a}\psi_{b\}}
{}^{cde})_0
\ea
\ee
where in $( \cdot )_0$ we removed the trace. Thus, the different
components in the differentials $d\varphi$ can be obtained from
\be972
\ba{rcl}
&
\ba{rcl}
\tau^{(1)} &\longleftrightarrow&  \psi \haken d\varphi \quad , \\
\tau^{(14)} &\longleftrightarrow& 
              \ast d\psi - {1 \over 4} (\ast d\psi) \haken\psi \ , 
\ea
& \qquad
\ba{rcl}
\tau^{(7)} &\longleftrightarrow&  \varphi \haken d\varphi \ , \\
\tau^{(27)} &\longleftrightarrow& 
              (d\varphi_{cde\{a}\psi_{b\}}{}^{cde})_0 \ ,
\ea
\ea
\ee
where $\tau_{14}$ and $\tau_{27}$ have to satisfy: $\varphi_3 \wedge
\Lambda_{27}^3= \varphi_3 \wedge \tau_{14} =0$.


\subsection*{SU(3) structures}


Having a $G=SU(3)$, one can find two singlet spinors on $X_7$, which
are equivalent to the existence of a vector field $v$.  This in turn
can be used to combine both spinors into one complex spinor defined as
\be726
\theta = {1 \over \sqrt{2}} \, e^Z \, ( \1 + v_a \gamma^a ) \theta_0
\quad , \qquad v_a v^a = 1
\ee
where the constant spinor $\theta_0$ is again the $G_2$ singlet and
$Z$ is now a complex function. The vector $v$ is globally well-defined
and gives a foliation of $X_7$ by a 6-manifold $X_6$ and both spinors,
$\theta$ and its complex conjugate $\theta^\star$, are chiral spinors
on $X_6$. In this case, we have to distinguish between the forms
$\Omega$ and $\tilde \Omega$ as defined in (\ref{624}) and
find\cite{110,130,470}
\be910
\ba{rcl}
\Omega^0 & = & e^{2 \, {\rm Re}(Z)} \ , \\[1mm]
\Omega^1 & = & e^{2\, {\rm Re}(Z)} \, v \ , \\[1mm]
\Omega^2 & = & i \, e^{ 2\, {\rm Re}(Z)} \, v\haken\varphi \, 
                = \, i \, e^{ 2\, {\rm Re}(Z)} \, \omega \ , \\[1mm]
\Omega^3 & = & i \, e^{ 2\, {\rm Re}(Z)} \, 
    \big[v\wedge ( v\haken\varphi)\big]
        \, = \, i \, e^{ 2\, {\rm Re}(Z)} \, v \wedge \omega \ ,\\[3mm] 
\tilde\Omega^3 & = & i \, e^{ 2\, {\rm Re}(Z)} 
   \big[ e^{2i\, {\rm Im}(Z)} \big( \varphi \, - \, v\wedge\omega 
            \, - \, i \, v\haken\psi \big) \big] \, 
   = i \, e^{ 2\, {\rm Re}(Z)} \, \Omega^{(3,0)} 
\ea
\ee
and all other forms are zero or dual to these ones.  The associated
2-form to the almost complex structure on $X_6$ is $\omega$ and with
the projectors ${1 \over 2} ( \1 \pm i \omega)$ we can introduce
(anti) holomorphic indices\footnote{Since the 6-d space is in general
not a complex manifold, we cannot introduce global holomorphic
quantities and this projection is justified only pointwise.}  so that
$\Omega^{(3,0)}$ can be identified as the holomorphic $(3,0)$-form on
$X_6$.  There exists a topological reduction from a $G_2$-structure to
a $SU(3)$-structure (even to a $SU(2)$-structure).  The difficulties
arise by formulating the geometrical reduction. Using the vector $v$,
let the explicit embedding of the given $SU(3)$-structure in the
$G_2$-structure be:
\be920
\ba{rcl}
\varphi & = & {\rm Re}( e^{-2 i\, {\rm Im}(Z)} \, 
\Omega^{(3,0)}) + v\wedge \omega \, =  \chi_+ + \, v \wedge \omega \ ,
\\[1mm]
\psi & = & {\rm Im}( e^{-2i\, {\rm Im}(Z)} \, \Omega^{(3,0)})\wedge v \, 
           + \, {1 \over 2} \omega^2 \, =\,  
\chi_- \wedge v \, + \, {1 \over 2} \omega^2
\ea
\ee
with the compatibility relations
\bea945
e^{- 2i\,  {\rm Im}(Z)} \, \Omega^{(3,0)} \wedge \omega & = & 
(\chi_+ \, + \, i \, \chi_-) \wedge \omega
            \, = \, 0 \, , \\
\chi_+ \, \wedge \, \chi_- & = & {2 \over 3} \, \omega^3\ .
\eea
Now, the projectors (\ref{512}) for $\theta_0$ imply for the complex
7-d in (\ref{726})
\[
\ba{rcl}
\gamma_a\theta &=& {e^Z \over \sqrt{2}} (\gamma_a + v_a 
       + i\varphi_{abc}v^b\gamma^c)\theta_0 \ , \\[2mm]
\gamma_{ab}\theta &=& {e^Z  \over \sqrt{2}}(i\varphi_{abc}\gamma^c
       +i\varphi_{abc}v^c + \psi_{abcd}v^c\gamma^d
       -2v_{[a}\gamma_{b]})\theta_0\ , \\[2mm]
\gamma_{abc}\theta &=& {e^Z  \over \sqrt{2}}(i\varphi_{abc} 
     + \psi_{abcd}\gamma^d +3iv_{[a}\varphi_{bc]d}\gamma^d
     - \psi_{abcd}v^d -4i\varphi_{[abc}\gamma_{d]}v^d)\theta_0\, , \\[2mm]
\gamma_{abcd}\theta &=& {e^Z  \over \sqrt{2}} (-\psi_{abcd} 
     - 4i \varphi_{[abc}\gamma_{d]} - 5\psi_{[abcd}\gamma_{e]}v^e \\
  && \qquad 
     - 4i v_{[a}\varphi_{bcd]} - 4v_{[a}\psi_{bcd]e}\gamma^e )\theta_0\ , 
     \\[1mm]
\gamma_{abcde}\theta &=& {e^Z  \over \sqrt{2}} (-5\psi_{[abcd}\gamma_{e]}
     -i\varepsilon_{abcdefg}\gamma^g v^f -5 v_{[a} \psi_{bcde]}
     -20i v_{[a} \varphi_{bcd}\gamma_{e]} )\theta_0\, , \\[2mm]
\gamma_{abcdef}\theta &=& {e^Z  \over \sqrt{2}} 
   (-i\varepsilon_{abcdefg}\gamma^g
     + \varepsilon_{abcdefg}v_h\gamma_j\varphi^{ghj} 
     -i \varepsilon_{abcdefg}v^g )\, \theta_0 \ .
\ea
\]
Again, these relations can be used to re-write the Killing spinor
equations in terms of constraint equations for the fluxes and a
differential equation for the warp factor as well as the spinor.  The
corresponding torsion components\cite{340} are now related to the
differential equation obeyed by the forms: $v$, $\omega$, $\Omega$ and
their dual.

As next case one would consider SP(2) structures implying three (real)
singlet spinors. An example is a 7-d 3-Sasaki-space (i.e.\ the cone
yields an 8-d Hyperk\"ahler space with Sp(2) holonomy), with the
Aloff-Walach space $N^{1,1}$ as the only regular examples\cite{360}
(apart from $S^7$); non-regular examples are in\cite{390}. We leave a
detailed discussion of this case for the future and investigate
instead the SU(2) case in more detail.


\subsection*{SU(2) structures}


On any 7-d spin manifold exist three no-where vanishing vector
fields\cite{380}, which implies that one can always define SU(2)
structures.  The corresponding four (real) spinors can be combined in
two complex SU(2) singlet spinors $\theta_{1/2}$.  The three vector
fields $v_\alpha$, $\alpha= 1,2,3$ can be chosen as
\[
v_1 \, = \, e^1 \qquad v_2\, = \, e^2  \qquad v_3 \, = \, \varphi(v_1,v_2)
\]
and they parameterize a fibration over a 4-d base space $X_4$. The
embedding of the SU(2) into the $G_2$ structures is then given by
\bea100 
\varphi &=& v_1\wedge v_2 \wedge v_3 + v_\alpha\wedge\omega_\alpha \ , \\
\psi &=& vol_4 + \epsilon^{\alpha\beta\gamma} v_\alpha\wedge v_\beta 
\wedge \omega_\gamma \ .
\eea
Since the vector fields are no-where vanishing, we can choose them of
unit norm and perpendicular to each other, i.e.\ $(v_\alpha , v_\beta)
= \delta_{\alpha\beta}$, and using the 3-form $\varphi$, one obtains a
cross product of these vectors.  One can pick one of these vectors,
say $v_3$, to define a foliation by a 6-manifold and on this
6-manifold one can introduce an almost complex structure by $J = v_3
\haken \varphi \ \in \ T^{\ast}M^6\otimes TM^6$.  The remaining two
vectors, which can be combined into a holomorphic
vector\footnote{Meaning, that it is annihilated by the projector: $(\1 -
J)$.} $v_1+ i\,v_2$ imply that this 6-manifold is a fibration over the
base $X_4$. On this 4-manifold we can define a basis of anti-selfdual
2-forms whose pullback correspond to the $\omega_\alpha$. Note, on any
general 4-d manifold we have the splitting
\[
\Lambda^2 = \Lambda^2_+ \oplus \Lambda^2_-
\]
where we can take $\{ \omega_1,\omega_2,\omega_3\}$ as a basis of
$\Lambda^2_-$ and this splitting appears in group theory as: $
\mathfrak{so}(4) \cong \mathfrak{su}(2) \oplus \mathfrak{su}(2)$.
The 2-forms satisfy the algebraic relations
\[
\omega_i^2 = 2 \, vol_4 \qquad 
               \omega_i\wedge\omega_j =0 \quad\mbox{for} \, i\neq j
\]
and the associating complex structures fulfill the quaternionic
algebra (note: the orientation on the 4-fold is negative).  We can
further split the 2-forms into a symplectic 2-form, say $\omega=
\omega_3$, and the remaining can be combined into complex
(2,0)-form. Thus, the subbundle $\Lambda^2_-$ decomposes as
\[
\Lambda^2_- \cong  \lambda^{2,0}  
         \oplus {\mathbb R} \, \omega 
\]
and besides the symplectic form $\omega$, we introduce the
complexified 2-form: 
\[
\lambda = \omega_1 + i\, \omega_2
\]
which is, with respect to $\omega$, a holomorphic (2,0)-form
(due to the quaternionic algebra satisfied by these forms). The
SU(2) singlet spinors can again be constructed from the $G_2$ singlet
spinor $\theta_0$ by
\be992
\theta_1 = \frac{1}{\sqrt{2}} (1 + v_3) \theta_0
\quad , \qquad
\theta_2 = v_1 \, \theta_1 
\ee
where $v_\alpha \equiv v^m_\alpha \gamma_m$. With the relations (\ref{512}),
it is straightforward to verify that: $(v_1 v_2 -i v_3) \theta_0 = 0$ and 
hence 
\[
(v_1 - i v_2) \theta_2 = (v_1 + i v_2) \theta_1 = 0 \qquad {\rm or:}
\quad
v_\alpha (\sigma^\alpha)_k{}^l \theta_l = \theta_k \  . 
\]
Moreover,
\be620
 \ba{rcl}
v_\alpha v_\beta \theta_k &=& \delta_{\alpha\beta} \theta_k
+ i \epsilon_{\alpha\beta\lambda} 
(\sigma^\lambda)_k{}^l \theta_l \ , \\[2mm]
\hat \omega \, \theta_k &=& 4 i \, \theta_k  \ , \\[2mm]  
\hat \lambda \, \theta_k  &=&  8 \, (\sigma_2)_k{}^l \theta_l^\star
\ea
\ee
where $\hat \omega \equiv \omega_{mn} \gamma^{mn}$, 
$\hat \lambda \equiv \lambda_{mn} \gamma^{mn}$ and
with the Pauli matrices
\be200
\sigma_1=\left( \begin{array}{cc} 0 & 1 \\ 1 & 0 \end{array} \right)
\ ,  \qquad
\sigma_2=\left( \begin{array}{cc} 0 & -i \\ i & 0 \end{array} \right)
\ , \qquad
\sigma_3=\left( \begin{array}{cc} 1 & 0 \\ 0 & -1 \end{array} \right)
\ .
\ee
For the forms (\ref{624}) we find
\be111
\ba{ll}
\Omega^{(0)} = \1 \ , & \tilde \Omega^{(0)} = 0\ ,\\[1mm]
\Omega^{(1)} = v^\alpha \sigma_\alpha \ , 
& \tilde \Omega^{(1)} = 0 \ ,\\[1mm]
\Omega^{(2)} = i \, \omega \, \1 + \Omega^{(1)} \wedge \Omega^{(1)} \ ,
\qquad
& \tilde \Omega^{(2)} = -  \lambda^\star \, \sigma_2 \ , \\[1mm]
\Omega^{(3)} =  \Omega^{(1)} \wedge \Omega^{(2)}  \ ,
& \tilde \Omega^{(3)} = - \Omega^{(1)} \wedge \tilde \Omega^{(2)} \ , \\[1mm]
\Omega^{(4)} = i\, \Omega^{(1)} \wedge \Omega^{(1)}
 \wedge \omega - vol_4 \, \1 \ , \ 
& \tilde \Omega^{(4)} = - \Omega^{(1)} \wedge \tilde \Omega^{(3)}  \ , \\[1mm]
\Omega^{(5)} =  \Omega^{(1)} \wedge \Omega^{(4)}  \ ,
& \tilde \Omega^{(5)} = \Omega^{(1)} \wedge \Omega^{(1)} \wedge \Omega^{(1)}
\wedge \lambda \sigma_2 .
\ea
\ee
%


\section{BPS constraints}


Now we can come back to the BPS equations from Section 2.  With the
superpotential as introduced before, equation (\ref{112}) becomes
\be827
0 = \tilde \eta
+ \big[\hat \gamma^5 \otimes \big( {1 \over 2} \partial A + 
{i m \over 36} \big)
+{1 \over 144} e^{-3 A}\, (\1 \otimes F) \big] \eta 
\ee
and if: $\hat \eta = e^{- {A\over 2} } \eta$, equation (\ref{711})
yields
\be553
0=  \1 \otimes
\big( \nabla_a^{(h)} +{i \, m \over 48} \, \gamma_a \big)\hat \eta
- i \,\hat \gamma^5 \gamma_a \,  e^{-{A \over 2}} \tilde \eta 
  - { 1 \over 12} \,  e^{-3 A}\, \hat \gamma^5 \otimes F_a  \hat \eta \ .
\ee
It is useful to decompose the 35 components of the 4-form field
strength under $G_2$ as {\bf 35} $\rightarrow$ {\bf 1 + 7 + 27} with
\be881
F_{abcd} ={1 \over 7} \, \F \, \psi_{abcd} 
          + \, {\mathcal F}^{(7)}_{[a} \, \varphi_{bcd]}
          - 2 \, {\mathcal F}^{(27)}_{e[a} \psi^e_{\ bcd]} 
\ee
where $\F$, $\FF$ and $\FFF$ are the projection introduced in
(\ref{971}).  The cases of $G_2$ and SU(3) structures have been
discussed already in the literature and we will summarize only the
main results.


\subsection*{$G_2$ structure}


In this case, the 11-d spinor is a direct product, i.e.
\be251
\eta = \epsilon \otimes \theta
\ee
and since the 11- and 7-d spinor are Majorana also the 4-d spinor
$\epsilon$ has to be Majorana (a more detailed discussion is
given in\cite{150}).  One finds that all internal 4-form
components have to vanish
\be311
F_{abcd} = 0  \quad , \qquad W_1 = 0 \  , \quad m = -36 W_2 \ .
\ee
The equation (\ref{553}) gives a differential equations for the spinor
$e^Z \theta_0$, which implies
\[
\partial_a Z =0 \ .
\]
The differential equations for $\theta_0$ fixes the 7-manifold to have
a weak $G_2$ holonomy and hence is Einstein with the cosmological
constant given by the Freud-Rubin parameter\cite{150,130}.  This in
turn implies, that the 8-d space built as a cone over this 7-manifold
has $Spin(7)$ holonomy. In fact, after taking into account the
vielbeine, this gives the known set of first order differential
equations for the spin connection 1-form $\omega^{ab}$: $\omega^{ab}
\varphi_{abc} = { 7\over 36} \, m \, e_b$, where $m$ was the
Freud-Rubin parameter [note $\omega$ is here the spin connection and
should not be confused with the associated 2-form introduced before].
Using the differential equation for the 7-spinor, it is
straightforward to verify that
\[
d\varphi =-{7 \, m \over 18} \, \psi \quad ,\qquad d \psi = 0
\]
and therefore only $\tau^{(1)}$ is non-zero.

The 4-d superpotential is only given by the Freud-Rubin parameter, ie.
\be883
 W^{(0)} \ \sim \ i \int_{X_7} {^{\star} F}
\ee
which fixes the overall size of the 7-manifold.  In the limit of flat
4-d Minkowski vacuum, the Freud-Rubin parameter has to vanish and we
get back to the Ricci-flat $G_2$-holonomy manifold. In order to allow
for non-trivial fluxes one has to consider SU(3) instead of $G_2$
structures.


\subsection*{SU(3) structure}


In this case, there is one (complex) 7-d spinor and the 11-d Majorana
spinor reads
\be010
\eta = \epsilon \otimes \theta + \epsilon^\star \otimes \theta^\star \ .
\ee
where the 4-d spinors $\epsilon$ and $\epsilon^\star$ have opposite
chirality ($\gamma^5 \epsilon = \epsilon$).  More details about this
case can be found in\cite{110,130}.  The solution of (\ref{827})
read now
\be524
\ba{rcl}
W= W_1 + i \, W_2 
&=& {1 \over 6} e^{-(K/2+3 A)} \,
  \big[ {4 \over 7} \F - v^a \FFF_{ab} v^b + i v^a \FF_a
  \big] \\[1mm]
v^a \partial_a e^{3A} &=&  {3 \over 7} \F + v^a \FFF_{ab} v^b 
\\[1mm]
m &=&0 \ .
\ea
\ee
and
\bea525
 (\delta_a^{\ b} - v_a v^b ) \FF_b & = &   
 \varphi_{abc} v^b \partial^c e^{3A} \ =\ 2 \, \varphi_a^{\ bc} v_b 
       \, \FFF_{cd} v^d 
\ , \\   \label{651}
2\, \FFF_{ab} v^b & = & \big[- {3 \over 7}  \F +  v^c \FFF_{cb} v^b
\, \big] v_a +  \partial_a e^{3A}
\eea
[the flux components were introduced in (\ref{881})].  In addition,
one obtains a differential equation for the spinor with the
non-trivial torsion components as introduced in (\ref{093})
\be761
\ba{rcl}
\tau^{(1)}  & \longleftrightarrow &  W_2 \ , \\
\tau^{(7)}_a \, &\longleftrightarrow& \, 48\, W_1 \, v_a
                        -{24 \over 7} \, {\mathcal F}^{(1)} \, v_a
          + \,  {3 \over 2} \, \varphi_{a}{}^{bc} v_b \FF_c
+ 27 \FFF_{ab} v^b \ .
\ea
\ee
To make the set of equations complete, we have to give the
differential equations obeyed by the vector field $v$, which is
straightforward if we use the differential equation for the
spinor
\be120
\ba{rcl}
\nabla_{m} v_{n} &=&  
- {1 \over 12} e^{-3A -2{\rm Re}(Z)}\, \theta^\dagger
 [ F_m , \gamma_n] \theta \\
&=&   {1 \over 12} 
e^{-3A} F_{mbcd} \omega^{bc} \omega^{d}_{\ n}
\ea
\ee
recall $\omega_{ab} = \varphi_{abc}v^c$. Note, $v^n \nabla_m v_n = 0$,
which is consistent with $|v|^2 =1$. Using the decomposition
(\ref{881}) one finds
\bea121
\nabla_{[m} v_{n]} &=& \big(\delta_{[m}^{\ a} \delta_{n]}^{\ b}
     +{1\over2}\psi_{mn}{}^{ab}\big) \FFF_{ac}v^c v_b 
              +{1\over4}\varphi_{mn}{}^a(\delta_a^{\ b} -v_a v^b) \FF_b
         \, , \\
\nabla_{\{m} v_{n\}} & = &  
      - {2 \over 7} (\delta_{mn} - v_{\{m} v_{n\}})\, \F
    -{1\over 2} v_{\{m}\varphi_{n\}}{}^{ab}v_a\FF_b  \nonumber \\ 
&&  +{1 \over 2}\big(
         \delta_m^{\ a} \delta_n^{\ b} + \omega_m^{\ a} \omega_n^{\ b}\big)
           \FFF_{ab}
      -{1 \over 2}\delta_{mn}\FFF_{ab} v^a v^b \ .   
 \eea
The first term in the anti-symmetric part is the projector onto the
${\bf 7}$, see (\ref{930}), and by contracting with $\varphi$ and
employing eqs.\ (\ref{525}) and (\ref{651}), one can verify
that\cite{110}: $d(e^{3A}v)=0$. One can project the flux components
onto $X_6$ and using the symplectic 2-form $\omega$ we can introduce
(anti) holomorphic indices. As result, we can define a 3-form $H$ and
4-form $G$ on $X_6$ and find for the superpotential
\be339
W = { i \over 36} \bar \Omega^{(0,3)} \haken H \qquad \rightarrow \qquad 
W^{(0)} \ \sim\   
{1 \over 36} \int_{X_7} F \wedge \Omega^{(3,0)} 
\ee
whereas the 4-form has to fulfill the constraint: $\Omega \haken G=0$
and $d e^{3A} \haken \omega = \frac{1}{2} \omega \haken H$ as well as
$v \haken d e^{3A} = \frac{1}{144} \omega^2 \haken G$.


\subsection*{SU(2) structure}


Finally, in the SU(2) case we write the 11-d spinor as
\be764
\eta = \epsilon^1 \otimes \theta_1 + \epsilon^2 \otimes
\theta_2 + cc 
\ee
and  we choose chiral 4-d spinors with 
\[
\hat \gamma^5 \epsilon^i = \epsilon^i \ .
\]
Eq.\  (\ref{827}) gives
\be521
0= \epsilon^i  \otimes \Bigl[ W_i{}^{j} \theta^\star_j
+ (\frac{1}{2} \partial A + \frac{im}{36} + \frac{1}{144}e^{-3A}  F ) \, 
\theta_i \, \Bigr]  \ .
\ee
If one does not impose any constraints on the spinors $\epsilon^i$,
one finds\cite{640}
\[
W_{ij} \sim \theta_i F \theta_j =  
       F \haken \tilde \Omega^{(4)}_{ij} 
\]
with the 4-form $\tilde\Omega^{(4)}$ as derived in (\ref{111}).
Defining the 2-forms: 
\[
G_{\alpha\beta} = v_\alpha^m v_\beta^n F_{mn ab} \lambda^{ab} 
\quad ,\qquad
F_{\alpha\beta} = v_\alpha^m v_\beta^n F_{mn ab} \omega^{ab}
\]
we can write $W_{ij}$ as matrix: $W \sim (\epsilon^{\alpha\beta\gamma}
G_{\alpha\beta} \sigma_\gamma) \sigma_2 $ with the $\sigma_\alpha$ as
Pauli matrices. It would be identical zero if: $G = 0$, but instead we
can also impose: $\epsilon^i W_{ij}=0$ so that $W_{ij}$ projects out
one of the 4-d spinor as we would need for an $\N = 1$ vacuum. This
implies that: $\det W =0$ which gives one (complex) constraint on the
2-form $G$. As next step, the contraction with $\theta^\dagger_k$
yields
\[
m = 0 \quad , \qquad dA \haken \Omega^{(1)} \sim F\haken \Omega^{(4)} 
\]
which implies that: $\partial_\alpha A \sim
\epsilon_{\alpha\beta\gamma} F^{\beta\gamma}$ (with $\partial_\alpha
\equiv v_\alpha^m \partial_m$) and $F\haken vol_4 = 0$.  Finally, one
has to contract with $\theta \gamma_a$ as well as with $\theta^\dagger
\gamma_a$ (with the index $a$ projected onto the base) and if we
assume that the $\partial_b A =0$ (ie.\ the warp factor is constant
over the 4-d base), we get as further contraints on the fluxes
\[
\theta^\dagger \gamma_a F \theta = 0 \quad , \qquad
\theta \gamma_a F \theta = 0 \ .
\]
These constraints are solved, e.g., if the only non-zero components of
$F$ are: $\sim v_\alpha \wedge v_\beta \wedge \omega$; ie.\ are
contained in $F_{\alpha\beta}$ and $G_{\alpha\beta}=0$ (as defined
above).

These are all constraints on the fluxes, but from the internal
variation (\ref{553}) we get differential equations. Setting, $m=0$
and $\tilde \eta=0$, we find
\[
\nabla_m \theta_i \ \sim \ F_{mnpq}\gamma^{npq} \theta_i 
\]
If only the components in $F_{\alpha\beta}$ are non-zero, it is
straightforward to further simplify this equation by using the
relations in (\ref{620}). On the other hand, this equation fixes also
the corresponding differential equations obeyed by the differential
forms. 
\[
\nabla_k \Omega^{(n)}_{m_1 m_2  \ldots }
 \sim F_k{}^{npq}\, \theta^\dagger [\gamma_{npq} , 
\gamma_{m_1 m_2 \ldots }] \theta \ .
\]
For the 2-forms eg., our constraints on the fluxes imply that
$\omega$ and $\lambda$ are closed, when projected onto the 4-d base,
which is therefore a hyper K\"ahler space.  Unfortunately, we have to
leave a detailed analysis of these equaions for the future.









\providecommand{\href}[2]{#2}\begingroup\raggedright\endgroup

\end{document}